# OPTIMIZATION OF SUPERCONDUCTING LINAC FOR PROTON IMPROVEMENT PLAN-II (PIP-II)*

Abhishek Pathak[†], Eduard Pozdeyev, Fermi National Accelerator Laboratory, [60510] Batavia, USA

*Abstract*

PIP-II is an essential upgrade of the Fermilab complex that will enable the world's most intense high-energy beam of neutrinos for the international Deep Underground Neutrino Experiment at LBNF and support a broad physics program at Fermilab. Ultimately, the PIP-II superconducting linac will be capable of accelerating the $H^-$ CW beam to 800 MeV with an average power of 1.6 MW. To operate the linac with such high power, beam losses and beam emittance growth must be tightly controlled. In this paper, we present the results of global optimization of the Linac options towards a robust and efficient physics design for the superconducting section of the PIP-II linac. We also investigate the impact of the nonlinear field of the dipole correctors on the beam quality and derive the requirement on the field quality using statistical analysis. Finally, we assess the need to correct the quadrupole focusing produced by Half Wave, and Single Spoke accelerating cavities. We assess the feasibility of controlling the beam coupling in the machine by changing the polarity of the field of linac focusing solenoids

## INTRODUCTION

The superconducting section of the PIP-II linac [1] aims to deliver a 5 mA, 800 MeV, $H^-$ beam with the asset of five distinct lineages of superconducting (SC) accelerating cavities accompanied by SC solenoids and normal conducting quadrupoles for transverse confinement. At such a high beam power, a meticulous, global optimization of the lattice parameters is indispensable to avoid transmission loss and beam quality deterioration via emittance growth, especially at lower energies where the particle dynamic is primarily driven by nonlinear space-charge forces. Here we present the results of a comprehensive lattice optimization study undertaken to ensure a reliable, efficient, and robust physics design through a stable region of operation in the stability chart [3] with an adiabatic variation in the phase advances while keeping the structure phase advances below $90^0$.

Apart from the beam degradation caused by the nonlinear space-charge forces at high intensities, the dipole corrector's nonlinearitys [5, 6] and asymmetric RF field produced by the spoke and HWR cavities [2, 4] have a deterministic role in helming the beam quality along the linac. Therefore, a detailed statistical analysis was performed to investigate the impact of the nonlinearity in the dipole corrector magnets on beam emittance, and an upper limit on the uniformity of the field integral was established for an acceptable beam quality throughout the linac. The HWR and SSR section of the linac operates with SC solenoids, guiding a symmetric focusing and, therefore, any divagation from a symmetric nature in the transverse (x-y) plane is disfavored. However, the central loading element used in these cavities introduces an asymmetric RF field, and therefore the beam suffers from asymmetric momentum impulses in x and y directions. The asymmetry produced by these cavities was investigated, and compensatory mechanisms using the solenoid current polarity and a quadrupolar field generated by the pairs of dipole corrector magnets with appropriate configurations were compared for efficient asymmetry minimization while introducing minimal coupling between the two transverse planes.

## LATTICE OPTIMIZATION

We considered the existing physics design of the PIP-II linac and performed optimizations to further enhance the linac efficiency and beam quality throughout the linac via a stable resonant-free operation. We performed a comprehensive investigation to quantify the impact of every accelerating and focusing structure on beam deterioration and instabilities along the linac and adopted a global tuning algorithm to minimize emittance growth and halo development in all three planes while maximizing the beam transmission. Here we determined to operate near a $k_x/k_{x,y}$ = 1.3, and therefore, the solenoid fields were adjusted to satisfy the chosen operating point in the Hofmann chart while keeping a smooth phase advance transition and the structure tune per period below $90^o$ as shown in Fig.1(a). As shown in Fig.1(b) (stability chart) and 1(c) (rms envelope evolution), the optimized physics design for the SC lattice of the PIP-II linac demonstrates a noteworthy resonance-free and stable operation with an emittance growth of 3.4% and 4.1% in transverse and longitudinal plane respectively, evading any transmission loss throughout the linac.

## CORRECTOR COIL NONLINEARITY AND EMITTANCE GROWTH

The HWR and SSR section of the PIP-II linac uses horizontal and vertical pairs of air-core dipole corrector magnets overlapped with the SC solenoids (Fig.2(a)) to correct for the misalignment-induced orbit oscillations in the lattice. The field uniformity within the region of interest for these corrector coils plays a significant role in governing the beam quality due to the build-up of sextupole terms with increasing nonlinearity. Fig.2(b) and 2(c) exhibit an azimuthal variation of field integral along R and the dominant Fourier coefficient for a dipole corrector with 11% field integral nonuniformity and exemplify the existence of sextupole field components. The existence of a sextupole component can cause emittance growth [5,6] and therefore needs to be minimized by estimating an upper bound on the field nonlinearity. We performed a detailed statistical study to model the cavities and solenoid misalignments within specified tolerances (Table-1) and followed a correction scheme using two corrector magnets and two BPSs to minimize the resulting orbit offsets. Here we used 500 random seeds to generate the misalignments following a Gaussian distribution and used dipole correctors with a nonuniformity percent of 5%, 7%, 9%, and 11% within a 12 mm radius to perform orbit corrections and examined the obtained transverse emittances at the exit of the linac.

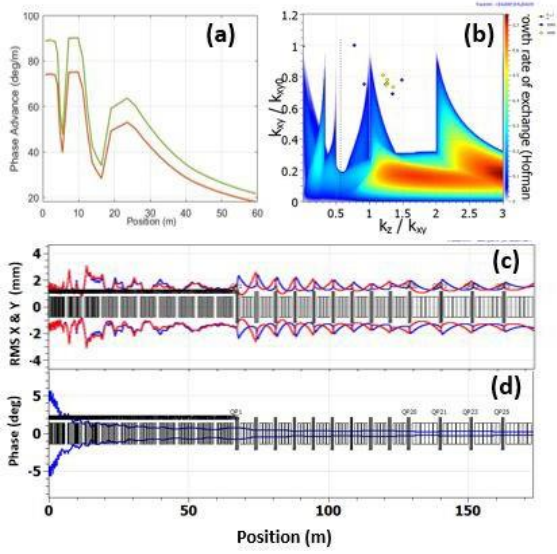

Figure 1: (a) Variation of structure phase-advance per period, (b) The stability chart overlapped with the period tune and phase advance ratios, and (c) the transverse and longitudinal envelope evolution along the linac.

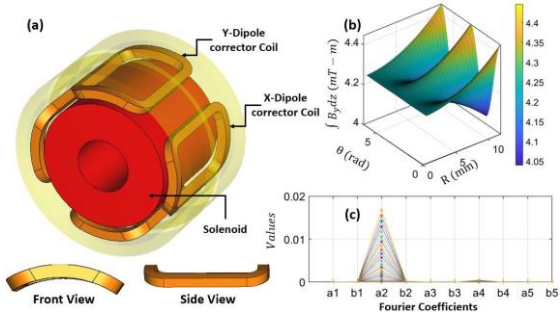

Figure 2: (a) Schematic of dipole corrector coils arrangement with solenoid and coil geometry, (b) Variation of field integral in R and $\theta$ plane, and (c) Fourier coefficient for varting radius from the centre.

Our study shows an increase in the rms percentage emittance growth from 2% to 6.3%, for a nonuniformity percentage increase from 5% to 11%, and the growth pursues a quadratic character as shown in Fig.3. However, the maximum percentage emittance growth reaches 6.5% and 20% for the mentioned percentage nonlinearities. Therefore a higher field integral nonuniformity can pose a grave peril to the beam quality throughout the linac. Following the trend shown by the emittance growth, we decided to keep an upper limit on the dipole corrector field integral nonlinearity within 5% so that the emittance growth can be restricted to a maximum value of 6.3% with an rms at 2%.

Table 1: Margin Specifications

| Parameters | X (mm) | Y (mm) | Z (mm) | Pitch (deg) | Yaw (deg) |
|---|---|---|---|---|---|
| HWR Solenoid | 0.5 | 0.5 | 1 | 1 | 1 |
| SSR-1 Solenoid | 0.5 | 0.5 | 1 | 1 | 1 |
| SSR-2 Solenoid | 0.5 | 0.5 | 1 | 1 | 1 |
| HWR Cavity | 0.5 | 0.5 | 1 | 3 | 3 |
| SSR-1 Cavity | 0.5 | 0.5 | 1 | 3 | 3 |
| SSR-2 Cavity | 0.5 | 0.5 | 1 | 3 | 3 |

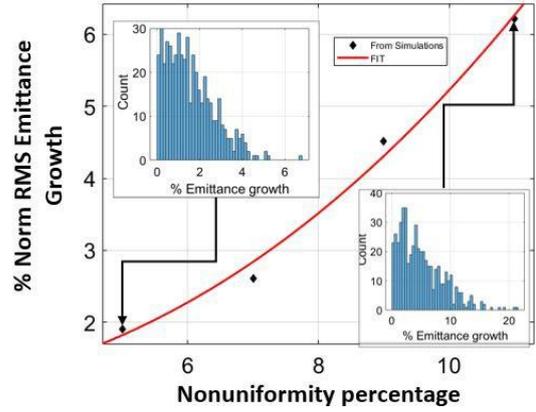

Figure 3: Variation in the RMS percentage growth with increasing field integral nonuniformity and % emittance growth distribution for 5% and 11% field nonuniformity.

## ASYMMETRIC RF KICK AND MITIGATION SCENARIO

Half wave resonators and spoke cavities are employed worldwide in hadron linacs at lower particle velocities because of their excellent efficiency and performance in the low $\beta$ range. Despite numerous advantages, the central loading element in such cavities introduces a transverse asymmetric RF field causing an elliptic transverse beam profile. Here we attempted to quantify the transverse splitting taking into account the transit-time factor and the synchronous phase of the beam, and explored methods to minimize it using the solenoid current polarities and the quadrupole fields generated by the corrector coils.

Our study shows that with a change in the current polarity of certain solenoids that are identified after considerable iterative optimizations, the x-y rms envelope splitting diminishes, as shown in Fig.4. But, a careful analysis of the beam matrix and the particle density distribution in coordinate space reveals that flipping the solenoid current polarity increases the x-y coupling and therefore the beam rotates with an elliptic profile giving a false impression of its symmetric nature.

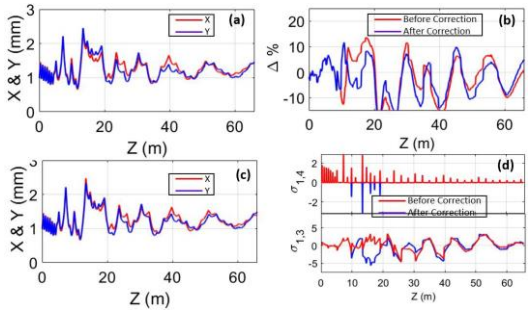

Figure 4: Variation in the RMS percentage growth with increasing field integral nonuniformity and % emittance growth distribution for 5% and 11% field nonuniformity.

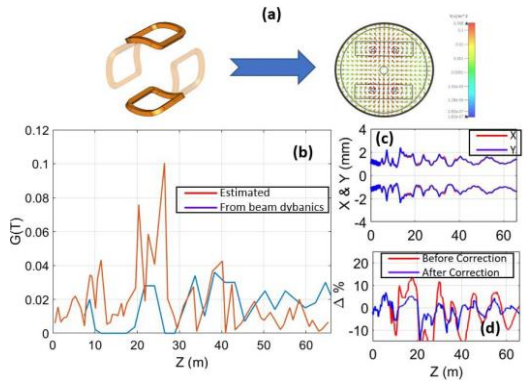

Figure 5: Variation in the RMS percentage growth with increasing field integral nonuniformity and % emittance growth distribution for 5% and 11% field nonuniformity.

With the limitations posed by the solenoid compensation approach, we examined the potential of utilizing the quadrupolar field produced by the corrector magnets with appropriate polarity. Fig.5(a) shows the pair of dipole corrector magnets and respective magnetic quadrupolar field profile. The x-y envelope splitting was calculated using the scaled RF fields from the optimized cavities, the synchronous phase, and particle velocity to estimate the compensating quadrupole gradient integral, and the results were compared with the compensating gradient integral obtained through a fully three-dimensional particle tracking simulations. Fig. 5(b) compares the calculated and obtained gradient integrals, which are in good agreement with each other. The use of the corrector magnets appears to reduce the splitting percentage from 15% to 5% in the SSR-1 section and from 13% to 10% in the SSR-2 section, as shown in Fig.5(c) and 5(d); however, the splitting remains within 5% for most of the linac. We calculated and compared the results before and after quadrupole corrections to establish the beam's transverse ellipticity reduction. and the results are plotted in Fig.6. Fig.6 shows and demonstrates an average reduction in the ellipticity from 0.17 to 0. 09 after quadrupole correction. Although we could achieve a significant reduction in the beam splitting, it exploits 50% of the correction strength; however, even without the quadrupole compensation, we observe a maximum ellipticity of 1.8, which has a negligible impact on the transverse emittance in the absence of other misalignments. Our study also shows that, in case of misalignment followed by orbit corrections, the splitting results remain unperturbed.

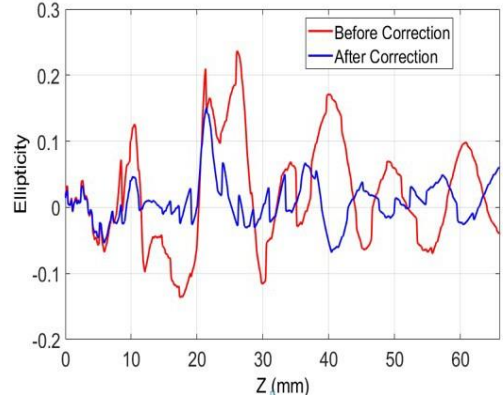

Figure 6: Beam ellipticity in X-Y plane before and after splitting minimization

## CONCLUSION

A comprehensive lattice optimization for the superconducting section of the PIP-II linac was performed to further enhance its stability and performance through a stable operation in a resonance-free region. The final lattice design with optimized phase advances and carefully tunned elements produces a nominal emittance growth of 3.4% and 4.1% in transverse and longitudinal planes, respectively while avoiding all resonance peaks in the stability chart.
We also performed a statistical study to investigate the effect of dipole corrector nonuniformity on emittance growth in the presence of cavity and solenoid misalignments followed by orbit corrections using the x and y pair of corrector coils. Our study indicates a quadratic dependence of emittance growth on the field integral nonuniformity, which we found to be acceptable within 5% for uncompromised beam quality. Finally, we looked at the beam splitting caused by the asymmetric RF fields from HWR and spoke cavities. After a detailed comparison of different splitting minimization approaches, the use of the quadrupole fields generated by the corrector coils is found to be most effective and reduces the beam ellipticity to a great extent but at the cost of utilizing 50% of the corrector strengths. Our analysis also suggests that the splitting compensation has a minimal effect on the beam qiality as the ellipticity even without compensation is nominal, and therefore the option of employing the dipole corrector coils as an asymmetry remedy may further be explored.